# A Survey on an Effective Defense Mechanism against Reactive Jamming Attacks in WSN

Manojkumar.M.K[1], Sathya.D[2]

[1]Student, Kumaraguru College of Technology, Coimbatore, Tamil Nadu, India.
[2]Assistant Professor, Department of Computer Science and Engineering, Kumaraguru College of Technology, Coimbatore, Tamil Nadu, India.

**ABSTRACT :** A Wireless Sensor Network (WSN) is a self-configure network of sensor nodes communicate among themselves using radio signals and deployed in quantity to sense, monitor and to understand the physical world. A jammer is an entity which interferes with the physical transmission and reception of wireless communications. Reactive jamming attack is a major security problem in the wireless sensor network. The reactive jammer stays quiet when the channel is idle. The jammer starts transmitting a radio signal as soon as it senses activity on the channel. The reactive jammer nodes will be deactivated by identifying all the trigger nodes, at the same time a jammer node is localized by exploiting the changes in the neighbor nodes. The affected node can be identified, by analyzing the changes in its communication range, compared to its neighbors. The paper proposes a survey on trigger node identification and a detailed survey on techniques to identify trigger nodes and highly concentrated on the reactive jammer.

## I. INTRODUCTION

A Wireless Sensor Network (WSN) is a self-configure ring network of sensor nodes communicating among themselves using radio signals and deployed in quantity to sense monitor and it understand the physical world. It consists of spatially distributed autonomous sensors to monitor environmental or physical conditions, such as temperature, vibration, sound, pressure, pollutants or motion and to cooperatively pass their data through the network to a main location.

The more modern wireless networks are bi-directional and also enabling to control the activity of the sensors. The improvement of wireless sensor networks was motivated by military applications such as battlefield surveillance, today such type of networks are used in many industrial and customer applications, such as industrial process observing and control, machine condition monitoring and so on.

### 1.1. Sensor Node

The WSN is built up of nodes, it consists of large number of sensor nodes, each sensor nodes are directly or indirectly connected to sink node. A wireless sensor node is composed of four basic components; there are sensing unit, processing unit (microcontroller), transceiver unit and power unit.

### 1.2. Sensing Unit

The main functionality of the sensing unit is to measure or sense the physical data from the target area. The signal or analog voltage is generated by the sensor corresponding to the observed phenomenon. The frequent waveform is digitized by an analog-to-digital converter (ADC) and then transported to the processing unit for further analysis.

### 1.3. Processing Unit

The processing unit is related with a small storage unit and it manages the processes that make the sensor nodes cooperate with the other nodes to carry out the assigned tasks.

### 1.4. Transceiver

There are three deploying communication schemes in sensors including optical communication (laser), radio-frequency (RF) and infrared (IF). Laser consumes less energy than radio-frequency and provides high security, but it requires line of sight and it is sensitive to atmospheric conditions. Infrared, laser, needs no antenna but it is limited in broadcasting capacity. Radio-frequency is the most easy to use but requires antenna. Various energy consumption reduction strategies have been developed such as filtering, modulation, and demodulation. Frequency and amplitude modulation are standard mechanisms. Amplitude modulation is simple but vulnerable to noise.

### 1.5. Power Unit

The power unit is one of the most important components of a sensor node. Every sensor node is armed with a battery that supplies power to remain in active mode. Power consumption is a main weakness of sensor networks. Any energy preservation schemes can help to extend sensor's lifetime. Batteries used in sensors can be categorized into two groups, rechargeable and non-rechargeable. In harsh environments, it is impossible to change or recharge a battery.

## II. JAMMING ATTACKS

A jammer is an entity who is purposefully trying to interfere with the physical transmission and reception of wireless communications [2]. A jammer continuously emits Radio Frequency signals to fill a wireless channel so that legitimate traffic will be completely blocked.

2.1. Jamming Attack Models

2.1.1. Constant Jammer

The Constant Jammer continuously emits a radio signal. It sends out random bits to the channel. It does not follow any MAC layer etiquette. It does not wait for the channel to become idle.

2.1.2. Deceptive Jammer

The Deceptive Jammer constantly injects regular packets to the channel. The normal nodes will be deceived by the packets. The normal nodes just check the preamble and remain silent. The jammer can only send out preambles.

2.1.3. Random Jammer

The Random Jammer alternates between jamming and sleeping. After jamming for $t_j$ unit of time, it turns off its radio and arrives sleeping mode. After sleeping for $t_s$ unit of time, it wakes up and resumes jamming constant or deceptive. The $t_j$ and $t_s$ may be fixed or random intervals-energy conservation.

2.1.4. Reactive Jammer

The Reactive Jammer stays quiet when the channel is idle. The jammer starts transmitting a radio signal as soon as it senses activity on the channel. It does not conserve energy because the jammer's radio must be continuously on in order to sense the channel and it is harder to detect.

Reactive jamming attacks have been considered as the most critical and fatally adversarial threats to subvert or disrupt the networks since they attack the broadcast nature of transmission mediums by injecting interfering signals. By identifying trigger nodes, the jammers can be avoided and completely nullify the reactive jamming attack.

## III. TECHNIQUES TO IDENTIFY TRIGGER NODES

The major techniques for identifying trigger nodes in Wireless Sensor Networks are as follows,
1) Non-adaptive Group Testing
2) Clique-Independent Set
3) Trigger-Node Identification

3.1.1. Non-adaptive Group Testing

The nature of our work is to identify all triggers in the large pool of victim nodes, so this technique is naturally matches our problem. The key idea of group testing is to test objects in multiple groups, instead of individually.

3.1.2. Traditional Non-adaptive Group Testing

The key idea of Traditional Non-adaptive group testing is to test items in multiple designated groups instead of testing them one by one[1][3]. The traditional Non-adaptive Group Testing method of grouping items is based on a designated 0-1 matrix $M_{t*n}$ where the matrix rows represent the testing group and each column refers to an item in Fig 1. M[i,j]=1 if the jth entry appears in the ith testing group, and 0 otherwise. The number of rows of the matrix represents the number of groups tested in Parallel [1]. Each entry of the result vector V refers to the test outcome of the corresponding group (row) where 1 represents positive outcome and 0 represents negative outcome.

$$M = \begin{pmatrix} 0 & 0 & 0 & 0 & 1 & 1 & 1 & 1 \\ 0 & 0 & 1 & 1 & 0 & 0 & 1 & 1 \\ 0 & 1 & 0 & 1 & 0 & 1 & 0 & 1 \\ 1 & 1 & 1 & 1 & 0 & 0 & 0 & 0 \\ 1 & 1 & 0 & 0 & 1 & 1 & 0 & 0 \\ 1 & 0 & 1 & 0 & 1 & 0 & 1 & 0 \end{pmatrix} \rightarrow V = \begin{pmatrix} 0 \\ 0 \\ 1 \\ 1 \\ 1 \\ 1 \end{pmatrix}$$

Fig 1. Matrix Table

The binary testing matrix M and the testing outcome vector V. Assumed that entry 1 first column and entry 2 second column are positive and only the first two groups are return the negative outcomes because they do not contain these two positive items. All the other four groups are return positive outcomes.

Given that, there are at most d < n positive entries among in total n ones, all the d positive entries can be correctly recognized on condition that the testing matrix M is d-disjunct matrix any single column is not contained by the union of any other d columns. Owing this property, each negative entry will appear in at least one row ie, group where all the positive items do not show up. By filtering all the entries appearing in groups with negative result, and all others are positive outcome. Although providing such simple decoding method, d-disjunct matrix is nontrivial which involve constructing with complicated computations with high overhead [3]. To improve this testing overhead, the deterministic d-disjunct matrix used to randomize error tolerant d-disjunct matrix, i.e., a matrix have less rows but remains the d-disjunct matrix. By introducing this matrix, the identification is able to handle test errors under sophisticated jamming environments.

### 3.1.3. Cliques-Independent Set

The Cliques-Independent Set is the problem to find a set of maximum number of pair wise vertex disjoint maximal cliques, which is referred as a maximum clique-independent set (MCIS)[4]. To the best knowledge, it has already been proved to be NP-hard for co comparability, line, planar, and total graphs; however, its hardness on UDG is still open. There have been numerous polynomial exact algorithms for solving the problem on graphs with detailed topology, e.g., Helly circular-arc graph and strongly chordal graph, but none of these algorithms gives the solution on UDG[4]. The scanning disk approach is used to find all maximal cliques on UDG, and find all the MCIS using a greedy algorithm.

## IV. TRIGGER NODE IDENTIFICATION

### 4.1. Anomaly Detection

Each sensor frequently sends a status report message to the base station. Once the jammers are activated during the time of message transmissions, the base station will not receive the reports from the sensors. By comparing the ratio of received reports with a predefined threshold value, the base station can decide whether a jamming attack is happened in the wireless sensor networks [1]. When generating the status report message, each sensor locally obtains jamming status and decides the value of the Label field. If a node v hears jamming signals, it will not send out messages but keep its label as victim. If v cannot sense jamming signals, then its report will be routed to base station, if it does not receive ACK from its neighbor on the next hop of the route within a time out period, it retransmits the signal. If no ACKs are received, it is possible that the neighbor node is a victim node, and then v updates Label tuple as boundary "BN" in its status report and another outgoing link from v with the available capacity is selected to forward this message. Base station with Label = TN will receive the status report and the corresponding node is regarded as unaffected and the messages are queued in the buffer of the intermediate nodes and forwarded in an first come first served manner. The TTL value is reduced by 1 per hop for each message, and any message will be dropped once it's TTL = 0.

The base station waits for the status report from each node in each period of length P. If no reports are received from a node v with maximum delay time, then v is considered as victim[1]. The maximum delay time is related to graph diameter. If the aggregate report amount is less than ψ, the base station starts to create the testing schedule for the trigger nodes in which the routing tables will be updated locally.

### 4.2. Jammer Property Estimation

The jamming range R and jammed areas are simple polygons based on the locations of the boundary and victim nodes. For sparse-jammer where the distribution of jammers is relatively sparse and there is at least one jammer whose jammed area does not overlap with the others[1]. By denoting the set of boundary nodes for the ith jammed area as $BN_i$, we can estimate the coordinate of this jammer as

$$(X_I, Y_I) = \left(\frac{\sum_{k=1}^{BN_i} X_k}{|BN_i|}, \frac{\sum_{k=1}^{BN_i} Y_k}{|BN_i|}\right) \quad (1)$$

where $(X_k, Y_k)$ is the coordinate of a node k is the jammed area and the jamming range R.

$$R = \min_{\forall BN_i} \left\{ \max_{k \in BN_i} (\sqrt{(X_k - X_J)^2 + (Y_k - Y_J)^2}) \right\} \quad (2)$$

for all the jammers have the same range.

For dense-jammer, jammed areas will be estimated based on simple polygons containing all the boundary nodes and victim nodes which consist of three steps which includes.

1) Discovery of convex hulls of the boundary and, does not contains unaffected nodes in the convex polygons.

2) For each boundary node v not on the hull, choose any two nodes on the hull and connect v to them in such a way that the internal angle at this reflex vertex is the smallest one and hence the polygon is modified by replacing an edge by the two new ones. The resulted polygon is the predictable jammed area [1].

3) Execute the near-linear algorithm to find the optimal variable-radii disk cover of all the victim nodes by execute the near-linear algorithm, but constrained in the polygon and return the largest disk radius as R.

### 4.3. Trigger Detection

If the jammer behavior is reactive, the straight forward way to find all the trigger node is, each sensor broadcast the signal one by one and listen to the possible jamming signals. The individual detection is quite time consuming and the entire victim nodes are isolated for a long detection period, or returns wrong detection result in the presence of mobile jammers.

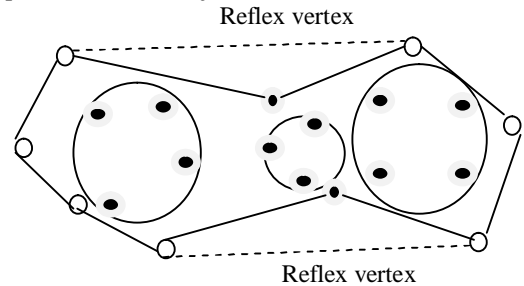

Fig. 2. Estimated R and jammed area

The performance of trigger detection protocol toward complicated attacker models with probabilistic attack strategies will be validated.

The base station designs the encrypted testing schedule over all the victim nodes based on the set of the global topology and the boundary nodes stored as a message as illustrated in Table 1 and all the boundary nodes are broadcasted. The broadcasting of the testing scheduling message adopts a routing mechanism similar to reverse path forwarding.

The nodes' IDs are recorded using their status report to the base station on their routing paths. Without considering the mobile jammers, the routing paths can be reused to send out these testing scheduling messages and evade the jammed areas.

TABLE 1 Trigger Detection Schedule

| Time Slot | Channel | Node List |
|---|---|---|
| 0 | $f_1$ | $v_1, v_3, \ldots v_n$ |
| 0 | $f_2$ | $v_1, v_2, v_4, \ldots v_{n-1}$ |
| 0 | : . | . . . . . . . . . . |
| 0 | $f_m$ | $v_2, v_5, v, \ldots v_n$ |
| 1 | $f_1$ | $v_2, v_4, \ldots v_{n-1}$ |
| : . | : . | . . . . . . . . . . |

After receiving the message, each boundary node broadcasts the message one time by using simple flooding method to its nearby jammed region. All the victim nodes execute the testing schedule and indicate themselves as non-triggers or triggers. All the sensor nodes are prepared with a global uniform clock and during the detection time, there is no message transmissions to the base station are required. The selection of the sets involves two-level grouping procedure.

First-level, all the set of victims are divided into several interference-free testing teams. The interference free means if the communications from the victim nodes in one testing team invoke a jammer node, its jamming region will not reach the victim nodes in another testing team. By trying broadcasting from victim nodes in each testing team and monitoring the jamming signals, the conclusion of any members in this team are triggers and all the tests in different testing teams can be executed simultaneously without interference with other testing team.

For example, three maximal cliques $C_1=\{v_1,v_2,v_3,v_4\}$, $C_2=\{v_3,v_4,v_5,v_6\}$, $C_3=\{v_5,v_7,v_8,v_9\}$ can be found within three jammed areas respectively in Fig 3. The three teams test at the same time. If $v_4$ in the middle team keeps broadcasting all the time and $J_2$ is awaken frequently, no matter the trigger $v_2$ in the leftmost team is broadcasting or not, $v_3$ will always hear the jamming signals, so these two are interfere with each other. In addition, node-disjoint groups are do not necessarily interference free as the rightmost and the leftmost teams.

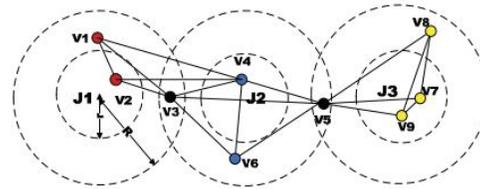

Fig. 3. Interference teams

Second-level, within the each testing team, victims nodes are divided into several testing groups. This is concluded by constructing a randomized (d,1) disjunct matrix mapping each sensor node to a matrix column and make each matrix row as a testing group (sensors corresponding to the columns with 1s in this row are chosen). The tests within one group will possibly interfere with another, so each group will be assigned with a different frequency channel for testing.

## V. CONCLUSION AND OPEN ISSUES

The survey deals with three techniques to identify trigger nodes and have identified some leftover problem.

Jammer mobility is one leftover problem to this framework, and the identification latency has been shown small, it is not efficient toward jammers that are moving at a high speed. Jammer localizations and jamming resistant routing are both quite promising, the service overhead has to be further reduced for real-time requirements.